\documentstyle[12pt,epsf]{article}

\begin{document}
\title{
TOPOLOGICAL PHENOMENA IN NORMAL METALS
\thanks{The main result of this work has been published
in ~\cite{novmal}. The present article is the extended lecture of authors
in the International Conference "Problems of Condensed Matter Theory" dedicated to the
80th birthday of I.M.Lifshitz, Moscow State University, June 1-4, 1997.
Its russian variant submitted in Uspekhi Phys. Nauk.}}
\author{ S.P. Novikov\thanks{Also at the Institute for Physical
Science and Technology, University of Maryland, College Park,
Maryland 20742-2431, e-mail: novikov@ipst.umd.edu}\,\, and A.Ya. Maltsev }
\date{ L.D. Landau Institute for Theoretical Physics,
 Ul. Kosygina 2,  Moscow 117940, Russian Federation,
e-mail: novikov@itp.ac.ru , maltsev@itp.ac.ru}
\maketitle

\vspace{0.5cm}

02.40.Vh, 72.15.Gd

\vspace{0.5cm}

\begin{center}
{\bf 1. Introduction. Historical remarks.}
\end{center}

Measurements of conductivity in the single crystal normal metals
in the strong magnetic fields lead to enormous number of different
phenomena. Many years ago (about 1956) I.M.Lifshitz, M.Ya.Azbel and
M.I.Kaganov formulated following principle:

All such halvanomagnetic phenomena should follow from the geometry
of semiclassical electron orbits, based on the one-particle Bloch
dispersion relations $\epsilon_{n}({\bf p})$ only ("Geometrical
strong magnetic field limit") - see ~\cite{lak}.

Essential new
features were found by I.M.Lifshitz and V.G.Peschansky
~\cite{lp1}-~\cite{lp2} - details see below. Important experimental
results were obtained in the works ~\cite{ag1}-~\cite{gaid}. Final
book describing results of this period was published
(see ~\cite{lifkag1}-~\cite{etm}). A lot of
concrete work was done since that, but general understanding
of this picture remained unchanged (~\cite{lifkag1}-~\cite{abr}).

About 1982 one of the present authors realized that this picture
leads to some beautiful problems of the low-dimensional topology
(~\cite{nov1}, see also ~\cite{nov2}-~\cite{nov4}). After that his pupils
A.Zorich, I.Dynnikov and S.Tsarev performed purely topological
investigations. They finally proved deep theorems, describing
topology of the generic open orbits and found geometrical
constructions for some very nontrivial nongeneric ("ergodic")
orbits (see ~\cite{zor}-~\cite{tsarev}).

Using these topological results, the present authors found universal
topological phenomena observable in the conductivity of single
crystal normal metals with "topologically complicated"
Fermi surface in strong magnetic field ~\cite{novmal} -
see below.

Let us point out that some noble metals have
topologically nontrivial Fermi surfaces. The first example of this
kind was Cu: its Fermi surface was found by Pippard in ~\cite{pippard};
Au, Pb, Pt, Ag belong also to this class. There are many other
examples now.

We concentrate our attention here on the results of ~\cite{novmal}
in the theory of normal metals. Using new topological results of
~\cite{dyn4}-~\cite{dyn6} it was observed that these ideas can
be applied also in the theory of semiconductors (details see in
~\cite{dynmal}).

\begin{center}
{\bf 2. Observable Quantities. Generic case.}
\end{center}

Let us consider any single crystal normal metal with lattice $L$
generated by the vectors ${\bf l}_{1}, {\bf l}_{2}, {\bf l}_{3}$.
As everybody knows, in the absence of magnetic field ${\bf B} = 0$
one-particle electron states can be described by the so-called
quasimomenta ${\bf p} = (p_{1}, p_{2}, p_{3})$ defined modulo
reciprocal lattice vectors:

\begin{center}
${\bf p}$ is equivalent to
${\bf p} + {\bf l^{*}}$
\end{center}

for any vector ${\bf l^{*}}$
such that $\langle {\bf l^{*}} , {\bf l}_{j} \rangle =
2 \pi \hbar n_{j}$, where every $n_{j}$ is an integer.
The reciprocal lattice $L^{*}$ is generated by the vectors
$({\bf l^{*}}_{1}, {\bf l^{*}}_{2}, {\bf l^{*}}_{3})$, such
that $\langle {\bf l^{*}}_{i} , {\bf l}_{j} \rangle =
2 \pi \hbar \delta_{ij}$.

In this approximation we have union of the "dispersion relations"
\begin{equation}
\label{disrel}
\epsilon_{n}({\bf p}) = \epsilon_{n}({\bf p}+{\bf l^{*}}),
\,\,\, n = 0,1,2,\dots ,
\end{equation}
describing dependence of the energy of electron on the quasimomenta.
In standard model electrons in the groundstate occupy all levels
below Fermi energy
$\epsilon_{F}:\,\, \epsilon_{j}({\bf p}) \leq \epsilon_{F}$;
All higher levels are empty. Conductivity theory in normal metals
deals with small perturbations of this picture. All essential
phenomena depends on the small neighborhood of the Fermi surface

Most known metals satisfy to the following nondegeneracy conditions:

a) There are no singular points of the dispersion relations on the
Fermi level:
$$\nabla \epsilon_{j}({\bf p}) \neq 0 \,\,\,\,\,
 for \,\,\,\,\, \epsilon = \epsilon_{F}.$$

b)\footnote{This property can be destroyed in some metals in strong
magnetic fields because of the magnetic breakthrough. See
explanations at the end of Chapter 3.}
Two different Fermi surfaces correspondent to different branches
$\epsilon_{j}({\bf p})$ and $\epsilon_{i}({\bf p})$ do not
intersect each other:
on the Fermi level $\epsilon = \epsilon_{F}$
$$\epsilon_{j}({\bf p}) \neq \epsilon_{i}({\bf p}), \,\,\, i \neq j$$

Let us describe the most important universal (under the nondegeneracy
conditions above) integer - valued observable topological quantities
for the conductivity in the strong magnetic fields following
results of the work ~\cite{novmal}. Apply strong magnetic field
${\bf B}$ (of the strength approximately $B \simeq 10 T$)
\footnote{As can be extracted from standard consideration of
these fenomena (see for example ~\cite{lifkag1}-~\cite{abr}),
there is the only restriction on the strength of magnetic
field: $\omega_{B} \tau \gg 1$, where $\omega_{B}$ is
Larmour frequency, $\tau$ - free electron motion time,
under which we can observe
our "geometrical limit". This condition leads to the
magnitudes of magnetic field $\sim 1T$ for pure gold samples
at temperatures $\sim 4 K$ used in the work ~\cite{gaid}.
Another restriction for the quasiclassical motion of electron is
$\hbar \omega_{B} \ll \epsilon_{F}$, but it is valid for all
admissible magnitudes of $B$ (the upper bound is of order of
$10^{3} \sim 10^{4} T$).}
and weak electric field ${\bf E}$ orthogonal to ${\bf B}$.
According to our results (~\cite{novmal}), there are two possibilities
only:

{\bf Case 1.} (Compact orbits)

2-dimensional part of conductivity tensor
$\sigma^{\alpha\beta}_{{\bf B}}$ tends to zero for $B \rightarrow \infty$,
${\bf B}/B$ fixed, (for $B \sim 10 T$ it is already very small):
$\sigma^{\alpha\beta}_{{\bf B}} \rightarrow 0, \alpha, \beta = 1,2$
in the plane orthogonal to ${\bf B}$.

{\bf Case 2.} (Generic open orbits)

For some direction ${\bf B}/B = {\bf n}$
the 2-dimensional part of conductivity tensor
$\sigma^{\alpha\beta}_{{\bf B}}, \alpha, \beta = 1,2$, tends to the
nonzero constant tensor $\sigma^{\alpha\beta}_{\infty}$ for
$B \rightarrow \infty$, depending on the direction of the unit
vector ${\bf n}$. In this case $(2\times 2)$-tensor
$\sigma^{\alpha\beta}_{\infty}$ has always rank equal to 1:
one of its eigenvalues is zero.
For the description of whole $3\times 3$ conductivity tensor
$\sigma^{ij}_{{\bf B}}, i,j = 1,2,3$ we introduce orthogonal
basis with first vector ${\bf e}_{1}$ directed along 0-eigenvector
of the $(2\times 2)$-tensor $\sigma^{\alpha\beta}_{{\bf B}}$ in the
plane orthogonal to ${\bf B}$, second vector ${\bf e}_{2}$ in the
same plane ${\bf e}_{2} \perp {\bf B}, {\bf e}_{2} \perp {\bf e}_{1}$,
${\bf e}_{3} = {\bf B}/B$ (see Fig.1).

\begin{figure}
\epsfxsize=4.0in
\centerline{
\epsffile{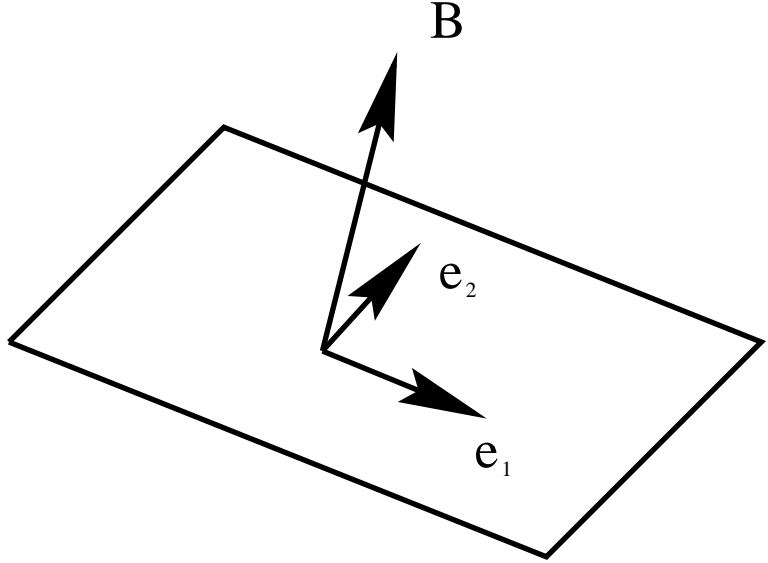}
}
\caption{Special basis corresponding to the Case 2.}
\end{figure}

We have

\begin{equation}
\label{sigmaij}
\sigma^{ij}_{{\bf B}} =
\left( \begin{array}{ccc}
0 & 0 & 0 \cr
0 & * & * \cr
0 & * & * \end{array}
\right) + O(B^{-1})
\end{equation}
where $(*)$ mean some nonzero constants.
In particular, $\sigma^{ij}_{{\bf B}} = \sigma^{ji}_{-{\bf B}}$
and $\sigma({\bf e}_{1}) = O(B^{-1})$.

This picture is locally stable: for the magnetic fields with
directions ${\bf e}^{\prime}_{3} = {\bf B}^{\prime}/B^{\prime}$
very closed to the original one ${\bf e}_{3} = {\bf B}/B$
we have conductivity tensor of the same structure as
(\ref{sigmaij}) in the new orthonormal basis
$({\bf e}^{\prime}_{1}, {\bf e}^{\prime}_{2},
{\bf e}^{\prime}_{3})$, where
$\sigma_{{\bf B}^{\prime}}({\bf e}^{\prime}_{1}) = 0, \,\,
\sigma^{\prime}_{\infty} \neq 0$ in the plane orthogonal
to ${\bf B}^{\prime}$.

Our most important statement is that the plane spanned by the
0-vectors ${\bf e}_{1}$ and ${\bf e}^{\prime}_{1}$ is integral;

\begin{figure}
\epsfxsize=4.0in
\centerline{
\epsffile{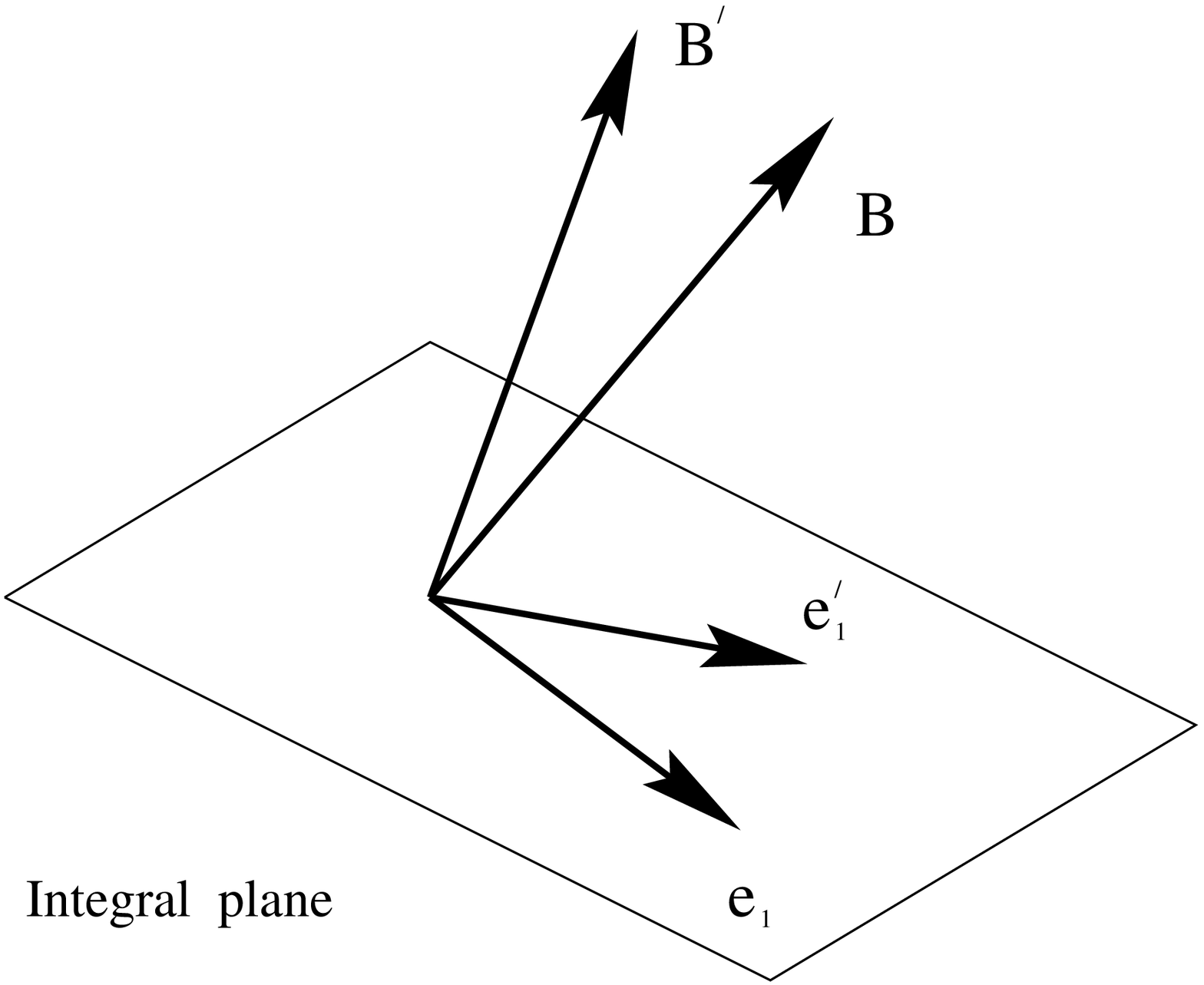}
}
\caption{Integral plane generated by vectors ${\bf e}_{1}$
and ${\bf e}_{1}^{\prime}$.}
\end{figure}

it is the same for all small rotations ${\bf B}^{\prime}$ of the
magnetic field ${\bf B}$.

Integrality means precisely that this plane is generated by two
reciprocal lattice vectors
$({\bar {\bf l^{*}}}, {\bar {\bar {\bf l^{*}}}})$:
$${\bar {\bf l^{*}}} = n_{1}{\bf l^{*}}_{1} +
n_{2}{\bf l^{*}}_{2} + n_{3}{\bf l^{*}}_{3}$$
$${\bar {\bar {\bf l^{*}}}} = m_{1}{\bf l^{*}}_{1} +
m_{2}{\bf l}^{*}_{2} + m_{3}{\bf l^{*}}_{3}$$
and
$${\bf e}_{1} = \alpha {\bar {\bf l^{*}}} +
\beta {\bar {\bar {\bf l^{*}}}} , \,\,\,
{\bf e}^{\prime}_{1} = \alpha^{\prime} {\bar {\bf l^{*}}} +
\beta^{\prime} {\bar {\bar {\bf l^{*}}}} .$$

Here $n_{j}, m_{j}$ - are the integer numbers. Components
of the vector ${\bar {\bf l^{*}}} \times {\bar {\bar {\bf l^{*}}}}$
characterize this plane. We are coming to triple of the integers:
$n_{1}m_{2} - m_{1}n_{2} = M_{3},\,\, n_{2}m_{3} - m_{2}n_{3} = M_{1},\,\,
n_{3}m_{1} - m_{3}n_{1} = M_{2}$; they defined up to the common
multiplier, so the invariant quantities are their ratios in fact.

We call this triple of integers $(M_{1}, M_{2}, M_{3})$
up to common multiplier the "Topological Type" of conductivity
tensor in strong magnetic field ${\bf B}$ defined by the pair
(or more) magnetic fields ${\bf B},{\bf B}^{\prime}$ with the
direction of ${\bf B}^{\prime}$ very close to ${\bf B}$ on
the unit sphere. Topological type is locally stable on the unit
sphere. After any small rotation of the magnetic field
topological type $(M_{1}, M_{2}, M_{3})$ remains unchanged.
Therefore it is the same on some open set of the directions on
the unit sphere. We call this open set a "Stability Zone" of
the type $(M_{1}, M_{2}, M_{3})$.

The area (or measure) of the Stability Zone of type
$(M_{1}, M_{2}, M_{3})$ on the unit sphere we shall denote
$\mu(M_{1}, M_{2}, M_{3})$. The area of the zone corresponding
to the Case 1 (above) we shall denote $\mu_{0}$.
We have following result

\begin{equation}
\label{result}
\mu_{0} + \sum_{(M_{1},M_{2},M_{3})}
\mu (M_{1}, M_{2}, M_{3}) = 4\pi
\end{equation}
(sum along all possible Topological Types).
For many Topological Types, in fact, we have:
$\mu (M_{1}, M_{2}, M_{3}) = 0$. Anyway, the Topological
Types with big enough integer numbers $|M_{j}|$ correspond
a very small value of this area $\mu$.

Therefore in real experiment we can observe only finite
(not big) number of Topological Types and their Stability Zones.
Mathematically, the equality (\ref{result}) means precisely
that all nongeneric possibilities (i.e. different from Cases 1 and 2)
correspond to the directions of ${\bf B}$ covering set
of zero measure on the unit sphere. Some especially interesting
nongeneric pictures will be discussed later.

For the comparison with old experimental data we give here
the asymptotic form of the resistance tensor, inverse to $\sigma$:\,\,
$R = \sigma^{-1}$ in the same basis as $\sigma$ above (\ref{sigmaij})
(see ~\cite{etm},~\cite{abr}).

{\bf Case 1.} Order of magnitude of ${\hat R}$ is:

\begin{equation}
\label{resis1}
{\hat R} \simeq {m^{*} \over ne^{2}\tau}
\left( \begin{array}{ccc}
1 & \omega_{B}\tau & 1 \cr
\omega_{B}\tau & 1 & 1 \cr
1 & 1 & 1 \end{array}
\right)
\end{equation}
(Part of matrix proportional to $B$ is skew-symmetrical).

{\bf Case 2.} Order of magnitude of ${\hat R}$ is:

\begin{equation}
\label{resis2}
{\hat R} \simeq {m^{*} \over ne^{2}\tau}
\left( \begin{array}{ccc}
(\omega_{B}\tau)^{2} & \omega_{B}\tau & \omega_{B}\tau \cr
\omega_{B}\tau & 1 & 1 \cr
\omega_{B}\tau & 1 & 1 \end{array}
\right)
\end{equation}
here $\omega_{B} = eB/m^{*}c$ - Larmour frequency,
and $\tau$ is free motion time of electrons.

Let us present here some experimental data obtained
by Yu.P.Gaidukov ~\cite{gaid} for Au. As we can see from
(\ref{resis2}), we should observe $B^{2}$ - dependence
of the resistance
$\rho \sim (B^{2} cos^{2}\alpha)\rho_{0}$
in the plane orthogonal to ${\bf B}$ in the Case 2.
Here $\rho_{0} = m^{*}/ne^{2}\tau$. The coefficient
$(cos^{2}\alpha)$ is equal to 1 for the electric field directed
along the vector ${\bf e}_{1}$ (above) - zero eigenvector
for conductivity tensor (\ref{sigmaij}) in the plane orthogonal to
${\bf B}$.

On the Fig.3 (Fig.11 in ~\cite{gaid}) we can see series
of black domains where $B^{2}$ - dependence has been observed
(1,0,0 means here the direction of vector {\bf B}, for example).

\begin{figure}
\centerline{
\epsffile{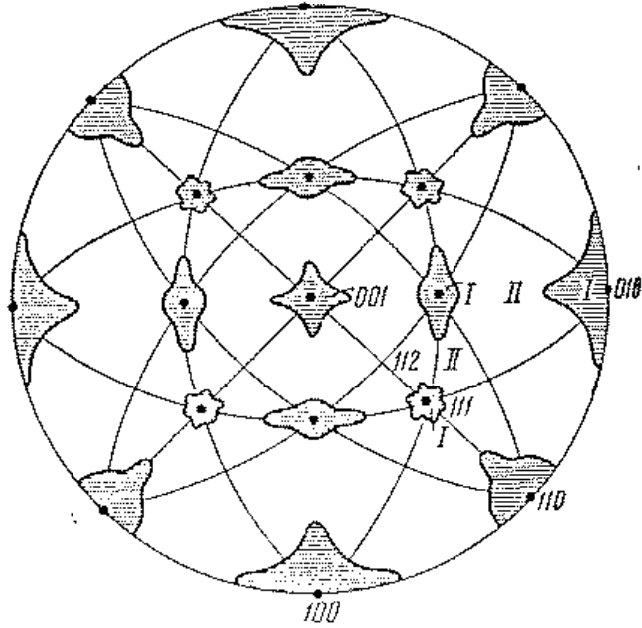}
}
\caption{Experimental data obtained by Yu.P.Gaidukov for Au. Black
domains correspond to Case 2. }
\end{figure}

It is interesting to point out that in the centers of the black
domains we find dots where resistance has "very deep minima"
and correspond to the Case 1, according to the results of ~\cite{gaid}.
The resistance within the black domains should be
$B^{2}$ - type, like in the Case 2. However, this dependence
was found experimentally like $B^{\alpha}$ for $\alpha \leq 2$
("slightly less", as written in ~\cite{gaid}). Probably,
magnetic field $B \sim 1 T$ was not enough for our asymptotic
behavior.
These experiments should be repeated with $B \geq 10 T$.
In the white area we have the Case 1. Something interesting
was found along the black lines (nongeneric situation ?).
We shall discuss it later.

Coming back to the interiors of black areas (with central dots removed),
we expect here final experimental confirmation of the
$B^{2}$ - dependence. In this case we can definitely state that
these black domains are really the "Stability Zones"
whose Topological Types correspond to the integral planes
orthogonal to unit vectors directed into the removed central
dots: so in the Fig.11 of ~\cite{gaid} - our Fig.3 -
we can see the following Topological Types:

$$(M_{1}, M_{2}, M_{3}) = (\pm 1, 0, 0), (0, \pm 1, 0),
(0, 0, \pm 1), (\pm 1, \pm 1, 0),$$
\begin{equation}
\label{mmm}
(\pm 1, 0, \pm 1), (0, \pm 1, \pm 1), (\pm 1, \pm 1, \pm 1).
\end{equation}
However, this statement should be verified experimentally.

Until now we consider the results of the important
work ~\cite{gaid} as a strong experimental evidence that
all black domains and some points on the black lines
do not belong to the Case 1. Our "integral planes" were not
known at that time: nobody asked about them.

Anyway, disappearance
of conductivity in the centers of black domains is in good agreement
with our understanding of this situation (Au):
if our integral plane is orthogonal to magnetic field ${\bf B}$,
and the generic open trajectories definitely exist for all magnetic
fields with directions closed to this one, they really can disappear
for this specific field ${\bf B}$ (see later).

{\bf Case 3.} Nongeneric situation (ergodic orbits).

Experimental results described in ~\cite{gaid} show some
strange behavior of the resistance tensor along black lines
on the Fig.3 : in many points of these lines resistance has very
deep maxima where asymptotic behavior like $B^{\alpha}$ has been
observed for different values of $\alpha$: $1 < \alpha < 1.8$.
In our opinion, these experimental results should be improved for
$B \geq 10 T$ instead of $B \sim 2 T$ like in ~\cite{gaid}.
We may expect here much more smaller stability zones of the generic
Case 2 with more complicated Topological Types
$(M_{1}, M_{2}, M_{3})$ then (\ref{mmm}), but there is
also another possibility:

We may expect here more complicated nongeneric "ergodic" orbits
of the types discovered in ~\cite{dyn4}-~\cite{tsarev}. By the
conjecture of the second author of this work this situation
"typically" leads to the scaling behavior of resistance
$R \sim B^{\alpha}, 1 < \alpha < 2$ (see ~\cite{maltsev}).

Anyway, some examples $(\gamma)$ of ergodic orbits give
such contribution in the asymptotic behavior of $3 \times 3$ -
conductivity tensor that:

\begin{equation}
\label{sigmerg}
\sigma^{ij}_{{\bf B}}(\gamma) \rightarrow 0 ,
\end{equation}
for $B \rightarrow \infty$ for all $i,j = 1,2,3$.

However we must add also contribution of all compact orbits. By the
geometrical reasons, we always have some compact orbits in the case
of Au (for any direction of magnetic field). Therefore we shall
always have nonzero strong magnetic field limit for the conductivity
component $\sigma^{zz}_{{\bf B}}$ along the magnetic field ${\bf B}$

$$\sigma^{zz}_{{\bf B}} \rightarrow \sigma^{zz}_{\infty} \neq 0 .$$

We expect to have much smaller value of $\sigma^{zz}_\infty$
for such nongeneric directions of ${\bf B}$ on the unit sphere
where ergodic orbits appear (in the comparison with generic
neighboring directions).
We should see local minima in such points of unit sphere.
The last property can be used to distinguish ergodic orbits
from small stability zones. There is no reasons to expect any
local minima of $\sigma^{zz}_{\infty}$ for small stability zones and
strong values of $B$ are needed for the observation of
$B^{2}$-dependence of resistance in $\Pi({\bf B})$.
More detailed discussion of the ergodic orbits see in~\cite{maltsev}.

\begin{center}
{\bf 3. Topological problems and explanations for the generic
case.}
\end{center}

We remind here that physical quasimomenta are presented by the vectors
${\bf p} = (p_{1}, p_{2}, p_{3})$ defined modulo reciprocal lattice
vectors. From the topological point of view such equivalence
classes can be considered as points in the 3 - torus $T^{3}$,
the "Brillouen Zone". All ${\bf p}$ - space $R^{3}$ we call
an "Extended Brillouen Zone".
For topologist it is a "universal
covering" over 3 - torus.

In the magnetic field ${\bf B}$ we use
standard semiclassical description of galvanomagnetic phenomena.
The "electron orbits" for the slow time evolution of the electron
Bloch waves can be obtained from dynamical system in
$({\bf x}, {\bf p})$ - space:
\begin{equation}
\label{xspace}
{\dot{\bf x}} = \{{\bf x}, \epsilon({\bf p})\}
\end{equation}
\begin{equation}
\label{pspace}
{\dot{\bf p}} = \{{\bf p}, \epsilon({\bf p})\}
\end{equation}
Here $\epsilon({\bf p})$ is the dispersion relation in the
absence of ${\bf B}$. Poisson brackets have following form:
$$\{p_{i}, x_{j}\} = \delta_{ij} , \,\, \{x_{i}, x_{j}\} = 0 , \,\,$$
\begin{equation}
\label{poisbr}
\{p_{i}, p_{j}\} = {e \over c} \,\, \epsilon_{ijk} \,\, B_{k}
\end{equation}

For the homogeneous magnetic field ${\bf B}$ our equations
(\ref{pspace}) for the variables $(p_{1}, p_{2}, p_{3})$ are
closed because $B_{k} = const$. We are coming to the hamiltonian
system on the 3 - torus (Brillouen Zone) with Poisson brackets
$$\{p_{i}, p_{j}\} = (e/c) \epsilon_{ijk} B_{k}$$
and Hamiltonian
$\epsilon({\bf p})$. It has 2 integrals of motion:
$\epsilon({\bf p})$ and $\sum B_{k}p_{k}$. The second integral
is a "Casimir" for this Bracket in the ${\bf p}$-space, because
$\epsilon_{jqk} = - \epsilon_{jkq}$:
$$\{p_{j}, \sum B_{q} p_{q} \} = e/c \sum_{q,k} \epsilon_{iqk}
B_{q} B_{k} = 0 $$

Therefore our electron orbits can be represented by 2 equations:
\begin{equation}
\label{orbits}
\epsilon({\bf p}) = \epsilon_{F} , \,\,\,
\sum B_{k} p_{k} = const .
\end{equation}
Geometrically, they are sections of the Fermi surface by the planes
orthogonal to magnetic field (every plane section is a union of
electron orbits).

We call electron orbit compact if it is closed
curve in the space $R^{3}$ (in the extended Brillouen Zone).
We call curve in $R^{3}$ periodic with period T and noncompact
if ${\bf p}(t+T) = {\bf p}(t) + {\bf l^{*}}$,
where ${\bf l^{*}}$ is some nonzero reciprocal lattice vector.
Strictly speaking, this curve is closed in the 3-torus $T^{3}$
(Brillouen Zone); topologist will say that such curve in the
3-torus is "nonhomotopic to zero". Compact curves are such that
${\bf l^{*}} = 0$. Topologist will say that they are homotopic to
zero in $T^{3}$.

We can easily see that periodic noncompact electron orbit may
appear only for the magnetic fields ${\bf B}$ such that the orthogonal
plane $\Pi({\bf B})$ contains at least one vector
${\bf l^{*}} \neq 0$, belonging to the reciprocal lattice.

Let us concentrate our attention on the generic "irrational"
magnetic fields ${\bf B}$ satisfying to the following restrictions:

1) The plane $\Pi({\bf B})$ does not contain reciprocal lattice
vectors.

2) All tangent points of $\Pi({\bf B})$ and Fermi surface are
nondegenerate. These points are critical points for the dynamical
system (\ref{pspace}) on the Fermi surface.

3)Separatrice of the saddle should never come to another saddle
for this dynamical system: it should have no second end at all
or should come back to the same saddle. Generically there is at most
one saddle on any plane parallel to $\Pi({\bf B})$ (orthogonal to
${\bf B}$).

Let us introduce a Topological rank of Fermi surface. The equation
$$\epsilon_{n}({\bf p}) = \epsilon_{F}$$
in the space of quasimomenta $R^{3}$ (Extended Brillouen Zone)
is given by the periodic function $\epsilon_{n}({\bf p})$.
This surface in the space $R^{3}$ is a union of "connected
components", on which any pair of points can be connected
by path on Fermi surface. We call Fermi surface "Topologically
complicated" if there is at least one connected component
(piece) of it which does not lie between any pair of parallel planes.
We say that such piece of Fermi surface has topological rank 3
(see Fig.4,a).

We say that Fermi surface has topological rank 2 if any its connected
component lies between some pair of parallel planes but there is one
which cannot be confined in any cylinder. This piece is like
"warped plane with holes" (Fig.4,b). It is possible that there is some
second piece (or more) of rank equal 2 with different direction
of parallel planes (see Fig.5).

Fermi surface has topological rank 1 if any its connected component
can be confined in some cylinder, but there is one which cannot
be confined in the box of finite size. This component is like
"warped cylinder" (Fig.4,c).

Fermi surface has rank 0 if any its connected component can be confined
in some box of finite size (Fig.4,d).

\begin{figure}
\epsfxsize=4.0in
\centerline{
\epsffile{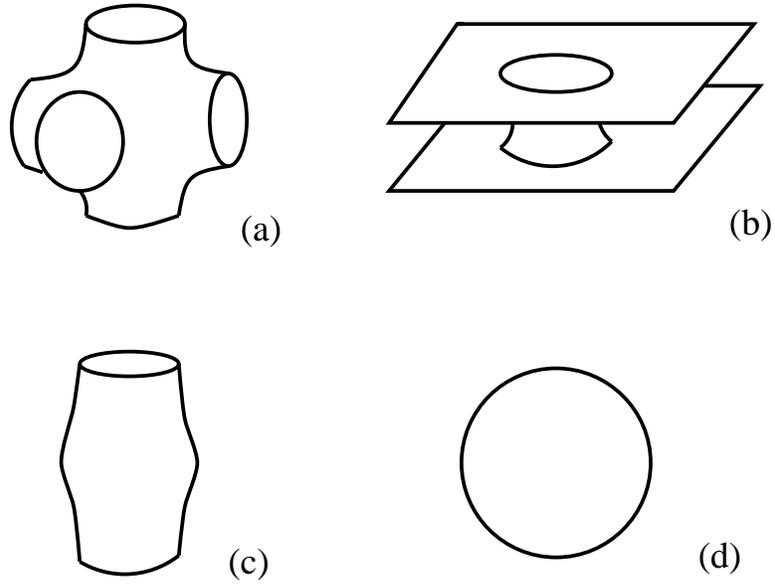}
}
\caption{Examples of Fermi surfaces of Topological Rank 3, 2, 1 and 0
respectively. }
\end{figure}

\begin{figure}
\epsfxsize=4.0in
\centerline{
\epsffile{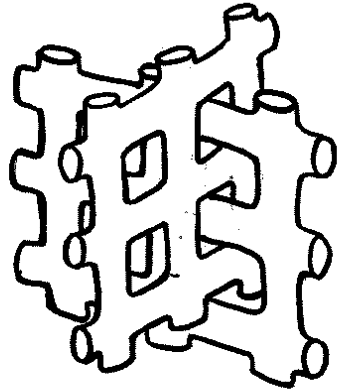}
}
\caption{Example of Fermi surface of Topological Rank  2
containing two components with different directions. }
\end{figure}

Applying magnetic field we get electron orbits on the Fermi surface
as the intersection of the planes $\Pi({\bf B})$ with it.
Following topological types are possible:

1. Fermi surface of the topological rank 0. All electron orbits
are compact.

2. Fermi surface of topological rank 1. Electron orbits can be
compact and open. Open orbits can exist only in the case when
magnetic field ${\bf B}$ is orthogonal to axis of the warped
cylinder on the corresponding piece. But even in this case
all orbits can be compact (for example for the "helix" -
see Fig.6).

\begin{figure}
\epsfxsize=4.0in
\centerline{
\epsffile{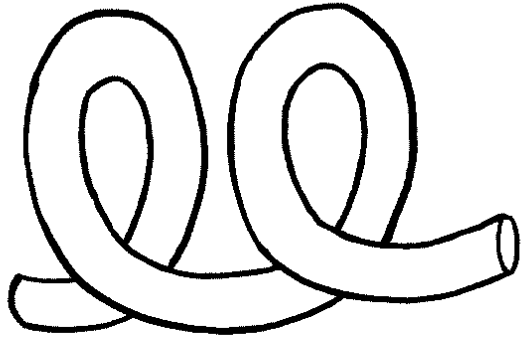}
}
\caption{Fermi surface of "helix" type. There is no open orbits
for any direction of ${\bf B}$.}
\end{figure}

 These open orbits (if they exist) should be
periodic with period-vector directed along the axis.
This picture is obviously nongeneric: open orbits may correspond
only to the 1-dimesional family of directions of ${\bf B}$
on the unit sphere.

3. Fermi surface of the topological rank 2. We can have compact
and open electron orbits here for any direction of magnetic field
${\bf B}$.

As was said above, generally speaking,
different connected components of our Fermi
surface may be confined between the parallel
planes with different (integral) directions. All open orbits,
if they exist, obviously have mean directions given by the intersection
of $\Pi({\bf B})$ with integral planes $\Pi_{j}(\epsilon_{F})$
of the corresponding connected
components. Integrality of these planes can be easily derived from
the periodicity of Fermi surface. If the direction
of ${\bf B}$ is irrational ($\Pi({\bf B})$ does not contains
reciprocal lattice vectors), open orbits can exist on the
components of rank 2 corresponding to one integral direction
only (we imply here that Fermi surface is nonselfintersecting).
Therefore all open trajectories have the same mean direction
given by the intersection of some integral plane $\Gamma$ with
$\Pi({\bf B})$. This picture is locally stable under small rotations of
${\bf B}$ and plane $\Gamma$ can be observed experimentally.

 This picture represents already the
situation corresponding to the generic Case 2 (above) for the
conductivity tensor. Topological Type here should correspond
to the very same integral planes $\Pi_{j}(\epsilon_{F})$, described
by the integers $(M^{j}_{1}, M^{j}_{2}, M^{j}_{3})$ for all ${\bf B}$
where open orbits exist (except the case
$\Pi_{j}({\bf B}) = \Pi(\epsilon_{F})$. For the nongeneric case
$\Pi({\bf B}) = \Pi(\epsilon_{F})$ all open orbits (if they exist)
are periodic in all components.

4. Consider now the most interesting case of Fermi surface with
Topological rank 3. We are dealing in fact with one piece of the
topological rank 3 only. Identifying the equivalent points
in the space of quasimomenta
$${\bf p} \equiv {\bf p} + {\bf l^{*}}$$
modulo reciprocal lattice vectors, we get a closed 2-dimensional
manifold (surface) in the 3-torus - "Brillouen Zone". Consider
now "fully irrational" magnetic field
${\bf B} = (B_{1}, B_{2}, B_{3})$ such that $\Pi({\bf B})$
does not contain integral (i.e. reciprocal lattice) vectors
${\bf l^{*}}$ and corresponding family of electron orbits
satisfies to the nondegeneracy conditions 1,2,3 (above).

Remove from Fermi surface all nonseparatrix compact orbits
(all of them are closed curves in the space of quasimomenta $R^{3}$).
Remaining part is obviously a union of surfaces, whose boundaries
are the separatrices:

Fermi surface
minus all nonsingular compact orbits = union of pieces $S_{i}$
(if there are open orbits). Boundary of $S_{i}$ is a set of some
closed separatrix curves $\gamma_{i\alpha}$ (see Fig.7)

\begin{figure}
\epsfxsize=4.0in
\centerline{
\epsffile{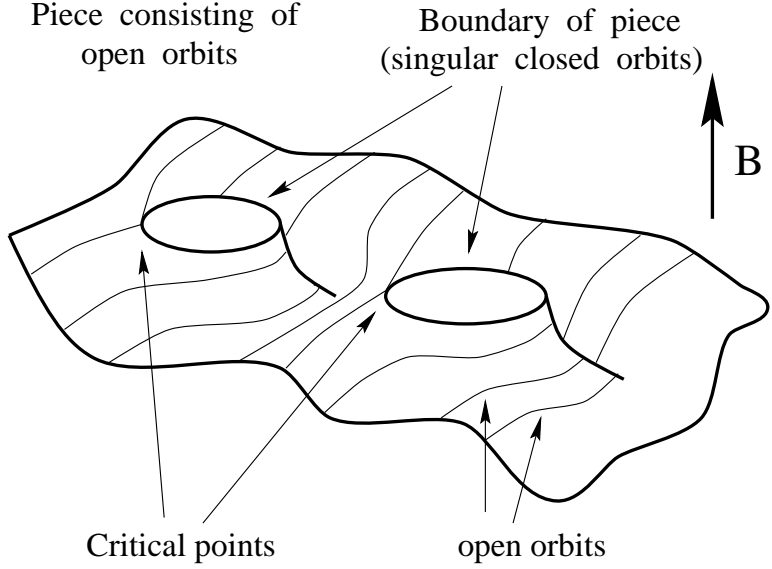}
}
\caption{Piece $S_{i}$ consisting of open orbits.
Singular closed orbits $\gamma_{i\alpha}$ with critical points are the
boundaries of $S_{i}$. }
\end{figure}

Every closed separatrix curve is a plane curve in $\Pi({\bf B})$.
Its interior is a topological 2-disc $D_{i\alpha}$. Fill in all
boundaries $\gamma_{i\alpha}$ by the plane 2-discs $D_{i\alpha}$
and add them to the surfaces $S_{i}$ (partial Fermi surfaces).
Finally we get closed 2-surfaces in $R^{3}$ (and their images
${\bar S}_{i}$ in the 3-torus $T^{3}$ after the identification
of the equivalent quasimomenta). By definition, all open orbits
lie on these pieces.

We call the genus of compact 2-manifold
${\bar S}_{i}$ in the 3-torus by the "genus of corresponding
open orbits", lying on it.

The most important nontrivial topological result, extracted by
the authors from the works of A.V.Zorich and I.A.Dynnikov
~\cite{zor}-~\cite{dyn6} is that generically all these pieces
${\bar S}_{i}$ have genus
1; it means that surfaces ${\bar S}_{i}$ are topologically
equivalent to the 2-dimensional tori, imbedded in the Brillouen Zone
- 3-torus $T^{3}$. The final general proof of this theorem is
very nontrivial (see ~\cite{dyn3}) and we shall not try to present
it here. "Generically" means in fact that if this statement
is wrong for some energy level
$\epsilon({\bf p}) = \epsilon_{0}$
than for all small nonzero perturbations of energy level
$\epsilon({\bf p}) = \epsilon_{1}$ it will be true.
We can easily find out that these surfaces ${\bar S}_{i}$
are in fact imbedded in $T^{3}$ without selfintersection.
They do not intersect each other as well. Last property
leads to the following result: every piece $S_{i}$ in
the space of quasimomenta $R^{3}$ looks like "warped plane"
after filling in by 2-discs $D_{i\alpha}$. Therefore its section
by the planes parallel to $\Pi({\bf B})$ lie in the strips of finite
width in these planes.

Local stability of this topology follows from the fact that all
compact curves in our constructions (like compact orbits and
compact separatrices with two ends in the same saddle, which is a
limit of compact nonsingular orbits) are stable under the small
rotations of the magnetic field. Our nondegeneracy requirements
for the metal and "generity" requirements for the direction of
${\bf B}$, are both important for this conclusion. From these
arguments follows full foundation of the Case 2 for the generic
conductivity tensor.

Let us point out that the first example of such topologically
stable open orbits was discovered by I.M.Lifshitz and
V.G.Peschansky in ~\cite{lp1} for the Fermi surface of Cu
(or for the so-called "thin" space net, as they called it
after Pippard's discovery of Fermi surface for Cu).
For this "thin space net" (see Fig.8 - Fig.2-3 in ~\cite{lp1})
we can have open orbits correspond to the Case 2 for the conductivity
tensor, where topological types may take only following values:
$$(M_{1}, M_{2}, M_{3}) = (\pm 1, 0, 0), (0, \pm 1, 0),
(0, 0, \pm 1).$$

\begin{figure}
\centerline{
\epsffile{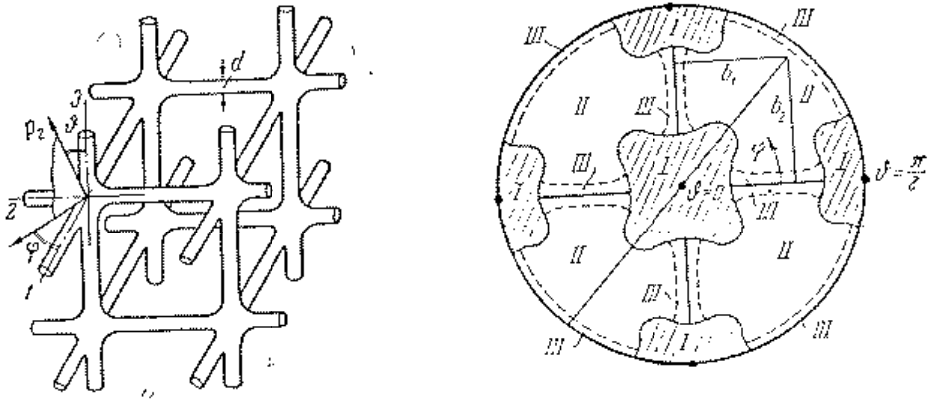}
}
\caption{The so-called "Thin space net" and stability zones
on the unit sphere corresponding to it. As was pointed by
I.M.Lifshitz and V.G.Peschansky, mean directions
of open orbits in these stability zones are given by the intersections
of $\Pi({\bf B})$ with coordinate planes (xy), (yz), (xz).}
\end{figure}

In the case of cubic lattice corresponding thin space net leads to the
stability zones of the small areas surrounding the directions above.
Other part of unit sphere belongs to the Case 1. It seems that
in the second work
~\cite{lp2} open orbits were found also for the more complicated
Fermi surfaces, but nothing like our integral planes was discussed.
The authors of ~\cite{lp2}, however,
thought that they found open domains on the unit sphere where
2 different mean directions of open orbits coexist (see fig.4 of
~\cite{lp2} - the last picture on our Fig.9).
This result of ~\cite{lp2} is mistakable and contradicts
to our results. This situation
is impossible for the open domains on the unit sphere. Our statement
is a mathematically rigorous corollary from the work ~\cite{dyn3}.

\begin{figure}
\centerline{
\epsffile{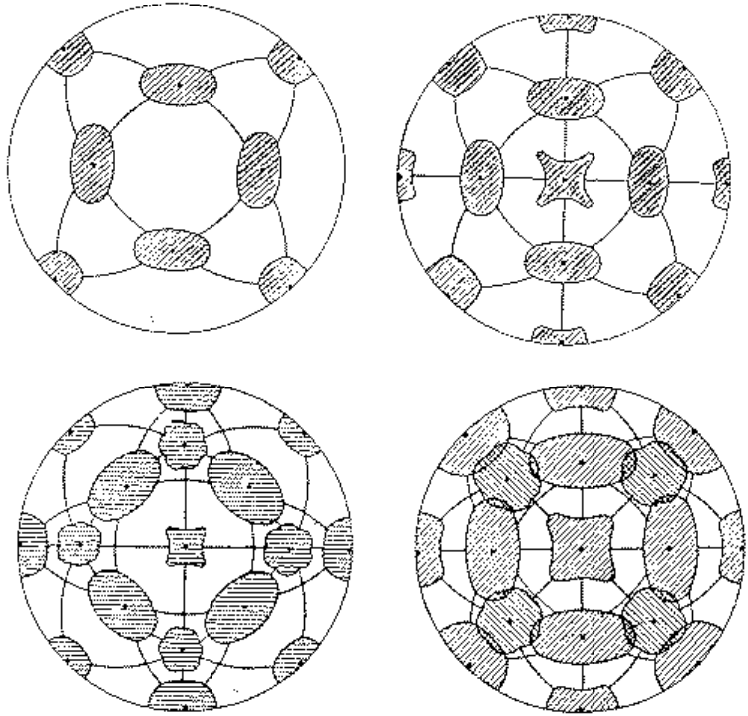}
}
\caption{Stability zones represented by I.M.Lifshitz and
V.G.Peschansky for different analytical examples of
Fermi surface for cubic lattice. For these zones our integral planes
were not discussed. According to our results,
the last picture is mistakable. Stability zones can not intersect
each other on the whole domain on the unit sphere and open trajectories
with different mean directions can not exist for the whole domain
of nonzero measure on the unit sphere. }
\end{figure}

In general, topological types of stability zones will be much
more complicated even for the same Fermi surfaces.

In the same works ~\cite{lp1}-~\cite{lp2} the contribution of the open
orbit ot this type in conductivity tensor was calculated. The last
result is valid for all open generic trajectories used in our paper
~\cite{novmal} and here. Besides that, our results are based on the
important additional property, extracted from topology:
All generic open orbits for given magnetic field have the same mean
direction.

The last property is true for the generic type of the open trajectories
only. For example, if magnetic field ${\bf B}$ is nongeneric
and $\Pi({\bf B})$ is an integral plane, than we may have open orbits
of different integral directions ${\bf l^{*}}$. In this case
every open trajectory is periodic. Everyone of them gives contribution
like in ~\cite{lp1}, but total sum of "partial conductivity tensors"
$\sigma^{\alpha\beta}_{i\infty}$ will have different algebraic structure
(for example, no zero eigenvector will be found in the plane
$\Pi({\bf B})$). Small rotation of ${\bf B}$ destroys this picture.
We can see that for the magnetic field ${\bf B}$ such that
$\Pi({\bf B})$ contains only one integral vector ${\bf l^{*}}$
(up to proportionality) such situation is impossible.

Let us point that it is possible for some symmetry
that two different Fermi surfaces (2 connected components
in mathematical terminology) can intersect each other in strong
magnetic field (see ~\cite{cohfal},~\cite{etm}).
It means precisely that electrons on each
component move completely ignoring other component in the limit
$B \rightarrow \infty$. Physical criteria for this phenomena
(magnetic breakthrough)
can be found in ~\cite{etm}. In this case only we can have in principle
two different families of open orbits with different mean directions,
such that this picture is stable under the rotations of ${\bf B}$.
Total sum of the contributions of both components in the conductivity
tensor will be a sum of the same types of tensors like in the generic
Case 2 with different directions of the integral planes. However
the Topological Types of partial integral planes have different
stability zones on the unit sphere. Therefore rotating magnetic
field we may observe their intersections, where we can see the case
more complicated then the Case 2. (M.I.Kaganov was the first who pointed
out to us on the magnetic breakthrough problem during the authors lecture.)

Consider now the nongeneric Case 3 of ergodic orbits. They
are unstable under the small rotation of the direction of ${\bf B}$
on the unit sphere. Let us remind that the Cases 1 and 2 are
"generic" in the sense that they correspond to the open everywhere dense
domain on the unit sphere. Using some additional
arguments (extracted from ~\cite{dyn6}) we are coming to the
conclusion that measure of the set of directions corresponding
to the ergodic orbits is equal to zero.
Its Hausdorf dimension is unknown.

The authors are grateful to M.I.Kaganov, V.G.Peschansky, L.A.Falkovsky
and Michael E. Fisher for their scientific help and advise.

\end{document}